\documentclass{mem}
\usepackage{natbib}
\usepackage{txfonts}
\usepackage{balance}
\usepackage{graphicx}
\usepackage[a4paper]{hyperref}
\idline{1}{1}

\begin{document}
\def\teff{$T\rm_{eff }$}
\def\kms{$\mathrm {km s}^{-1}$}
\input{psfig}

\title{
Cosmological implications of Gamma Ray Bursts
}
   \subtitle{}

\author{
Gabriele Ghisellini\inst{1} 
          }

  \offprints{G. Ghisellini}

\institute{
Istituto Nazionale di Astrofisica --
Oss. Astron. di Brera, Via Bianchi, 46
I--23806 Merate, Italy
\email{gabriele.ghisellini@brera.astro.it}
}

\authorrunning{G. Ghisellini}

\titlerunning{Cosmology with Gamma Ray Bursts}

\abstract{ 
The discovery that the bolometric energetics (and/or peak luminosity)
of Gamma Ray Bursts correlates with their spectral properties
has allowed to standardize the burst energetics to such a degree
to enable their use for constraining the cosmological parameters,
in the same way as SN Ia. 
With respect to SN Ia, there is the advantage of having
sources free from extinction problems, and easily detectable
also at large redshifts.
On the other hand, these spectral--energy correlations are not yet
understood, and bursts with a complete set of information
(to standardize their energetics) are still few (two dozens).
There have been already attempts to use these bursts to
constrain $\Omega_\Lambda$ and $\Omega_{\rm M}$, and even
the dark energy equation of state.
These results are very encouraging.

}
\maketitle{}

\section{Introduction}

Gamma Ray Bursts (GRBs) are powerful.
We can compare their emitted power with the Planck power,
i.e. the Planck energy divided by the Planck time, which can also be
written as
\begin{equation}
L_{\rm P} \, =\, { Mc^2 \over R_{\rm g}/c} \, =\, {c^5 \over G}
\, \sim 3.6\times 10^{59}\,\,\, {\rm erg \, s^{-1}}
\end{equation}
i.e. a mass entirely converted into energy in a time equal to the light
crossing time of its gravitational radius $R_{\rm g}$ 
($G$ is the gravitational constant).
GRBs can emit, in electromagnetic form, $L\sim 10^{52}$--$10^{53}$
erg s$^{-1}$, while Active Galactic Nuclei can have
luminosities up to $10^{48}$ erg s$^{-1}$ (but for a much longer time),
and Supernovae can have $L\sim 10^{43}$ erg s$^{-1}$ for a month, and
$L\sim 10^{45}$ erg s$^{-1}$
for a few hundreds seconds during the shock breakout.
Due to their power, even relatively modest $\gamma$--ray instruments
have no difficulties in detecting them also at high redshifts.
Furthermore, hard X--rays can travel unabsorbed
across the universe: with their largest power and least
absorption, GRBs are thus ideal candidates to study the far universe.

\section{Standard candles?}

The energetics of the prompt emission of GRBs 
span at least four orders of magnitudes: at first sight,
GRBs are all but standard candles.
However, there are a few correlations between the total bolometric
energetics and the spectral properties of bursts which can be used
to standardize the GRB energetics.
In general, ``blue" GRBs (having the peak of their prompt spectrum 
at higher energies) are more powerful/energetic (contrast this with 
blazars, behaving exactly the opposite way; Fossati et al. 1998).
These correlations are named after the discoverer, and in the
following I try to summarize them.

{\bf Frail: universal energy reservoir? ---}
Frail at al. (2001, see also Bloom et al. 2003)
found that the collimation corrected energetics
of those GRBs of known jet aperture angles
clustered into a narrow distribution, hinting to
a ``universal energy reservoir" 
$E_{\gamma} =(1-\cos\theta_{\rm j}) E_{\rm \gamma, iso} \sim 10^{51}$ erg.
The aperture angle of the jet is estimated in the following way.
Consider a shell moving with a bulk Lorentz factor $\Gamma$.
Unlike blazars, the motion is radial, not unidirectional.
Due to aberration, the observer will see only a fraction $1/\Gamma^2$ 
of the emitting surface. 
But $\Gamma$, during the afterglow, is decreasing.
At some time $t_{\rm j}$, the fraction of the observed surface 
becomes unity. This happens when $\Gamma = 1/\theta_{\rm j}$.
Before $t_{\rm j}$ the increased fraction of the observable surface
partially compensates for the decreasing emissivity,
while after $t_{\rm j}$ this compensating effect ends.
Therefore one expects a break in the light curve at $t_{\rm j}$.
Since only geometry is involved, this break should be achromatic
(Rhoads 1997).
Knowing the dynamics of the system (i.e. how $\Gamma$ changed in time),
we can derive $\theta_{\rm j}$.
The dynamics is controlled by the conservation of energy and momentum,
leading to the self similar law $M_{\rm ISM} = m_{F}/\Gamma = 
E_{\rm F} / (\Gamma^2 c^2)$, where $M_{\rm ISM}$ is the mass swept by the fireball
at a given time, $\Gamma$ is the bulk Lorentz factor at that time,
and $E_{\rm F}$ is the energy of the fireball.
If the process is adiabatic, the latter is constant.
With this law, we obtain $\theta_{\rm j} = \Gamma(t_{\rm j})^{-1}$ and then
\begin{eqnarray}
\theta_{\rm j} &=& 0.161 \,
\left({ t_{\rm jet,d} \over 1+z}\right)^{3/8} 
\left({n \, \eta_{\gamma}\over E_{\rm iso,52}}\right)^{1/8};
\,\,\,  \quad {\rm H} \nonumber \\
\theta_{\rm j} &=& 0.2016 \,
\left( {t_{\rm jet, d} \over 1+z}\right)^{1/4} 
\left( { \eta_\gamma\ A_* \over E_{\rm iso,52}}\right)^{1/4}; \quad {\rm W } 
\label{theta} 
\end{eqnarray}
where $n$ is the circumburst density in the homogeneous (H) case,
$z$ is the redshift and
$t_{\rm j,d}$ is the break time measured in days.
The efficiency $\eta_\gamma$ relates the
isotropic kinetic energy of the fireball 
$E_{\rm k, iso}$ to the prompt emitted energy $E_{\rm iso}$: 
$E_{\rm k, iso}= E_{\rm iso}/\eta_\gamma$. 
Usually, one assumes a constant value
for all bursts, i.e. $\eta_\gamma =0.2$ (after its first use by Frail et al.
2001, following the estimate of this parameter in GRB 970508; Frail et al. 2000).

For the wind (W) case,  $n(r)=Ar^{-2}$ and $A_*$ is the value of 
$A$ [$A=\dot M_{\rm w} /(4\pi v_{\rm w})=5\times 10^{11}A_*$ g cm$^{-1}$] 
when setting the wind mass
loss rate to $\dot M_{\rm w} =10^{-5} M_\odot$ yr$^{-1}$ and the
wind velocity to $v_{\rm w}=10^3$ km s$^{-1}$. 
Usually, a constant value (i.e. $A_*=1$) is adopted for all bursts.

\begin{figure*}[t!]
\psfig{figure=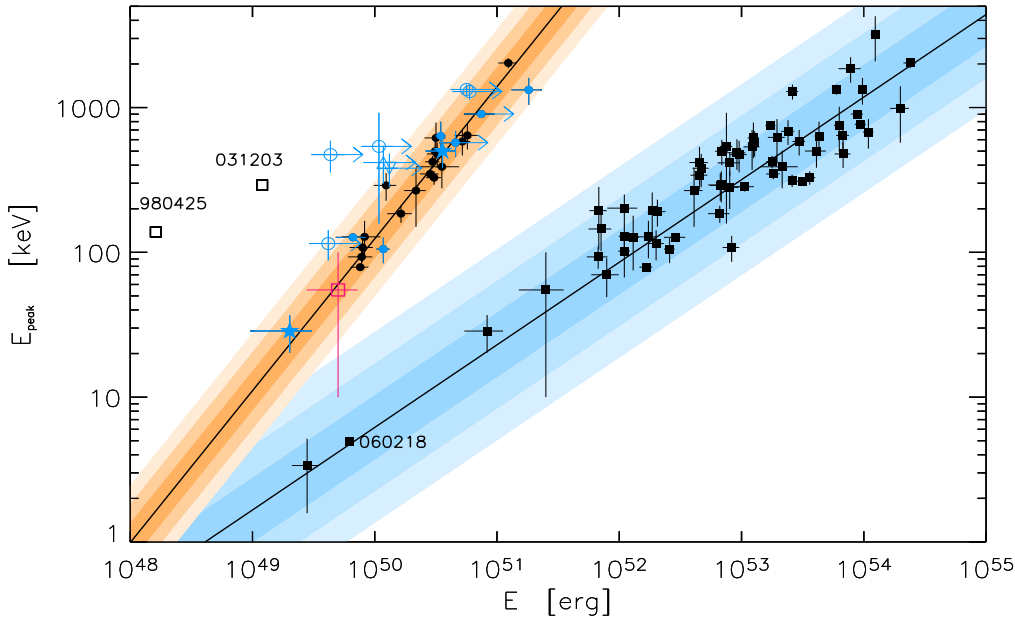,width=14cm,height=9cm}
% \resizebox{\hsize}{!}
% {\includegraphics[clip=true]{wind_ama.ps}}
\caption{
% \footnotesize
The most updated (Jan. 2007) 
Amati and the Ghirlanda correlations.
The latter is shown in the case of a circumburst
material with a wind density profile and contains
25 objects. The red point is...
The number of objects for the Amati correlation is 62.
(From Ghirlanda et al. 2007a). As can be seen, apart from the two
anomalous GRBs (980425 and 031203), there are no new outliers.
The solid lines are the best fits, and the three shaded regions represent the
regions of 1, 2, 3 $\sigma$ scatter around the best fits.
The empty square is GRB 060614, not included in the fit.
}
\label{amawind}
\end{figure*}
\begin{figure*}[t!]
% \resizebox{\hsize}{!}
% {\includegraphics[clip=true]{lz05.ps}}
\psfig{figure=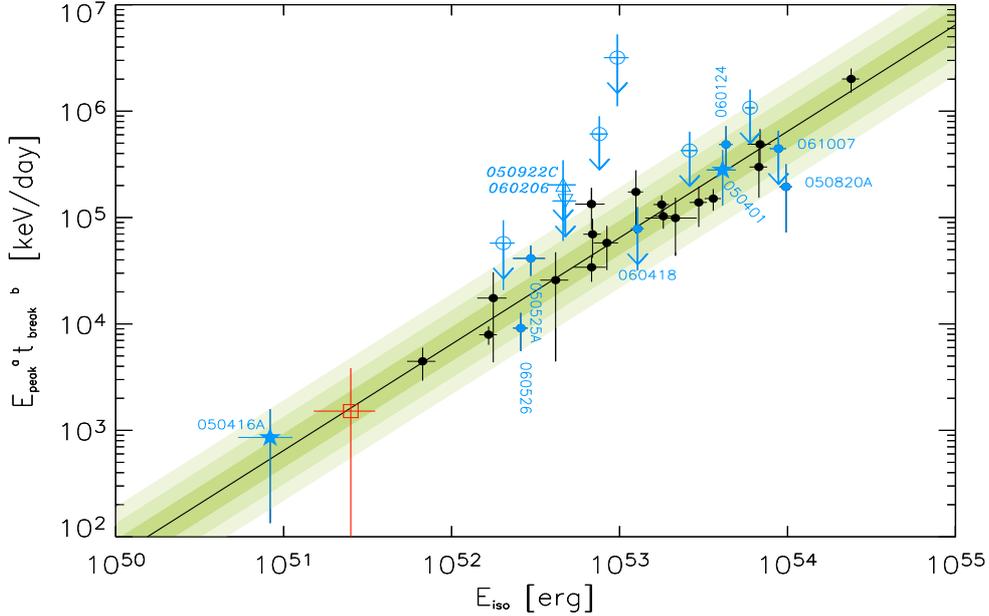,width=14cm,height=9cm}
\caption{
\footnotesize
The most updated (Jan. 2007) Liang \& Zhang correlation.
The best fit exponents are $a=1.88\pm0.15$ and $b=0.92\pm 0.13$.
Symbols as in Fig. \ref{amawind}
(From Ghirlanda et al. 2007a).
}
\label{lz05}
\end{figure*}
\begin{figure*}[t!]
\begin{tabular}{cc}
% \resizebox{\hsize}{!}
% {\includegraphics[clip=true]{firmani.ps}}
\psfig{figure=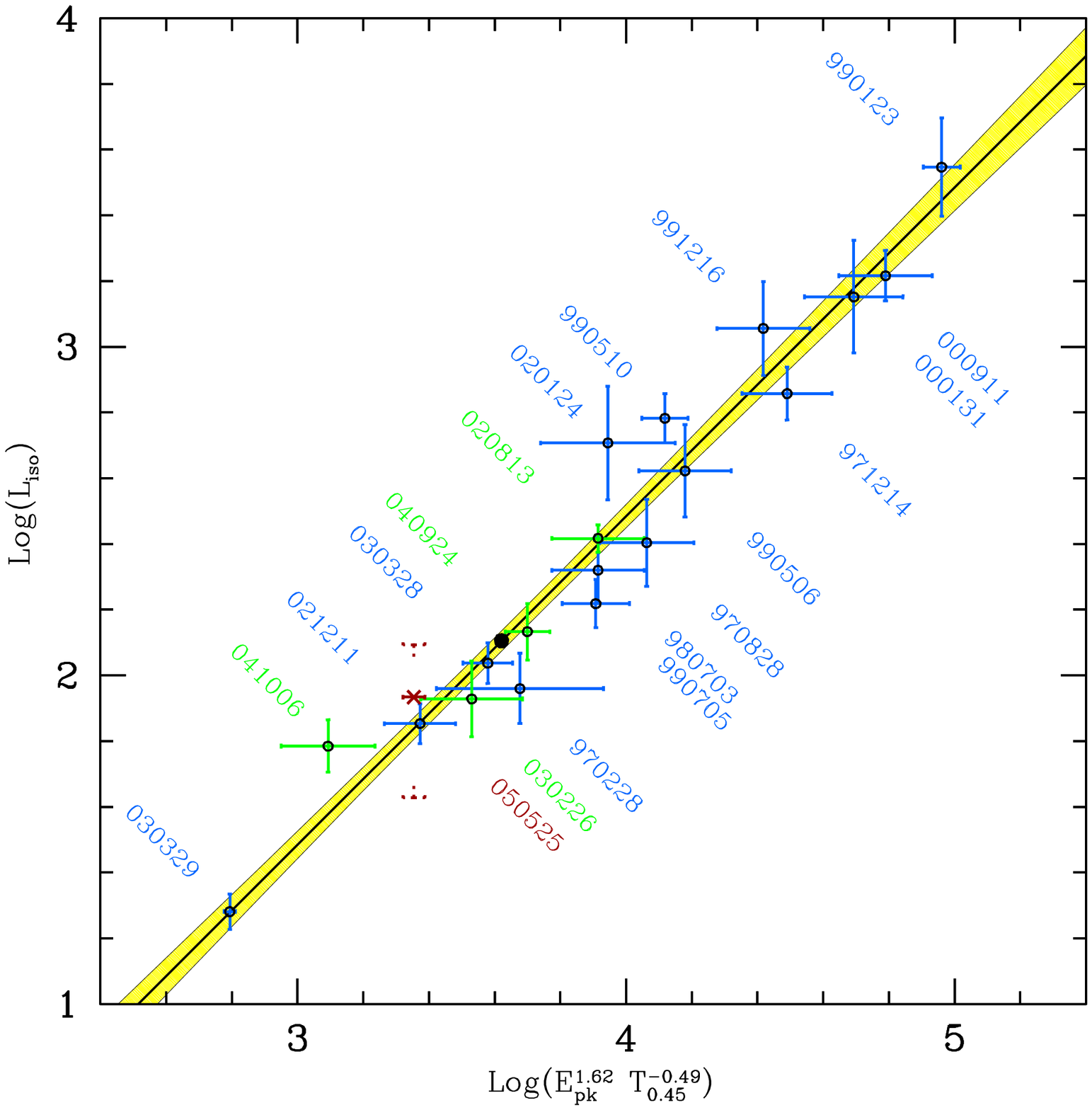,width=7cm,height=7cm}
&\psfig{figure=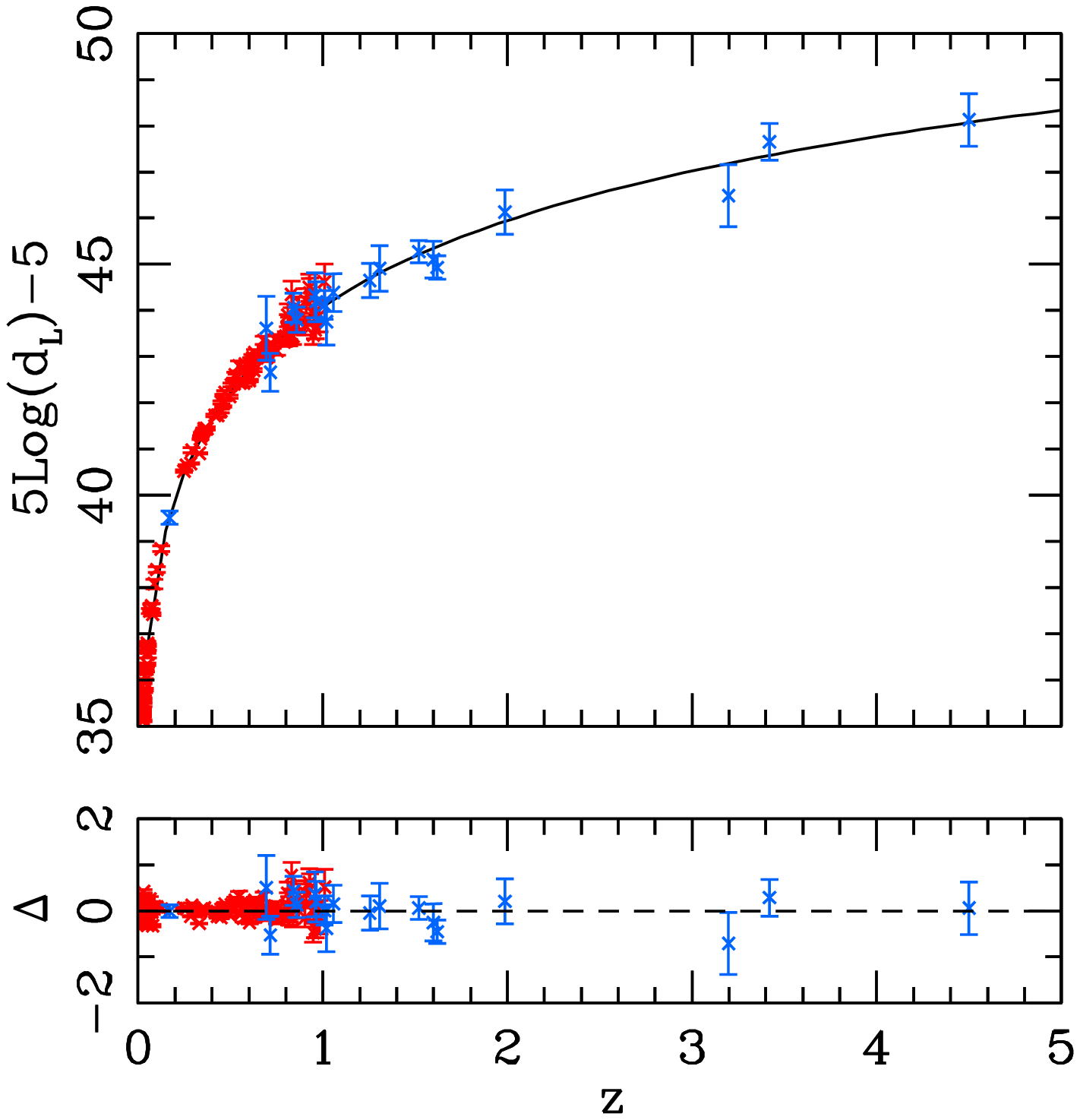,width=7cm,height=8cm}
\end{tabular}
\caption{
\footnotesize
Left: The Firmani correlation. Right: Hubble diagram obtained
using GBRs and SN Ia together. The light grey contours refer to
SN Ia alone. GRBs have been standardized using the Firmani correlation,
SN Ia comes from the sample of Astier (2006). The bottom panel
shows the residuals.
(From Firmani et al. 2006a and 2006b). 
}
\label{f06}
\end{figure*}

{\bf The Amati correlation ---}
Amati et al. (2002), considering $Beppo$SAX bursts, 
found that the isotropic energetics
correlate with the peak energy $E_{\rm p}$ of the time integrated
prompt emission: $E_{\rm p} \propto E_{\rm iso}^{1/2}$.
This correlation, expanded in later works (Amati 2006, Ghirlanda et al. 2007),
is obeyed by all but two bursts (the anomalous GRB 980425 and GRB 031203,
but see Ghisellini et al. 2006) for which the redshift and $E_{\rm p}$
is known. 
Claims by Nakar \& Piran 2005 and Band \& Preece 2005
that the Amati correlation is spurious, resulting from selection
effects, were contrasted by Ghirlanda et al. (2005), using
a large sample of GRBs for which the pseudo--redshift were derived
by the lag--luminosity relation.
Fig. \ref{amawind} show the updated (Jan. 2007) Amati correlation
which includes 62 GRBs (plus the two outliers).

{\bf The Yonetoku correlation ---}
Also the peak luminosity $L_{\rm p, iso}$ of the prompt emission
correlates with $E_{\rm p}$, in the same way as $E_{\rm iso}$:
$E_{\rm p} \propto L_{\rm p, iso}^{1/2}$ (Yonetoku et al. 2004).
The scatter is similar to the scatter of the Amati correlation.
Since the luminosity $\propto \Gamma^2$, this correlation
has the same form also in the comoving frame, contrary to
the Amati one.

{\bf The Ghirlanda correlation ---}
By correcting the isotropic energetics by the factor $(1-\cos\theta_{\rm j}$),
Ghirlanda et al. (2004) found that collimation corrected energy, $E_{\gamma}$,
is not universal, but is tightly correlated with $E_{\rm p}$.
To find $\theta_{\rm j}$, Eq. \ref{theta} for the homogeneous case
was originally used, with $t_{\rm j}$ derived from the optical light curves.
The efficiency $\eta$ was assumed to be constant, as well as the density of the
interstellar medium (unless it was derived by means, in a very few cases).
The correlation is $E_{\rm p} \propto E_{\gamma}^{0.7}$.
Later, Nava et al. (2006) considered a wind density profile and
an updated list of GRBs (18 objects), and found a linear correlation:
$E_{\rm p} \propto E_{\gamma}$.
The linear form is particularly intriguing for two reasons.
First, it means that it has the same linear form also in the
comoving frame, since $E_\gamma$ and $E_{\rm p}$, being two energies, transform
in the same way.
The second reason is that $E_\gamma/E_{\rm p}$ is constant.
This ratio is the number of photons at the peak, which must be the same
for all bursts and it is approximately $10^{57}$ (coincidentally,
the number of protons in a solar mass).
The most updated correlation, using 25 GRBs and including Swift bursts
(Ghirlanda et al. 2007a), confirms the earlier results, as can be
seen in Fig. \ref{amawind} (wind case).

{\bf The Liang \& Zhang correlation ---}
To find the jet angle, one needs a model, which in turn requires to know
the efficiency and the circumburst density and profile.
The model, based on energy and momentum conservation, appears robust,
while the assumption of the same efficiency and same density
for all bursts is questionable.
On the other hand, the fact that the angle resulting from this assumption
allows to construct a very tight correlation is an indication that the
distribution of values of the efficiency and of the density must be
narrow, or, alternatively, that $(\eta_\gamma n)$ is a function of $E_{\rm p}$.
If not, the tightness of the correlation is fortuitous.
These concerns are by--passed by the existence of
the Liang \& Zhang (2005) correlation, which is entirely phenomenological,
i.e. it is model independent and assumption--free.
It involves three observables (plus the redshift) and it is of the form
$E_{\rm iso} \propto E_{\rm p}^2 t_{\rm j}^{-1}$.
In Nava et al. (2006) we have shown that if the exponent of $t_{\rm j}$ 
is close to unity (as it is), then the Liang \& Zhang correlation is entirely
consistent with the Ghirlanda correlation.
The tightness of the Liang \& Zhang correlation is similar to the Ghirlanda one.
Note that $t_{\rm j}\propto E_{\rm iso}^{-1}$ for burst with the
same $E_{\rm p}$ (this is the reason of the clustering found by Frail
et al. 2001, since their bursts had similar $E_{\rm p}$; see Nava et al.
2007).

{\bf The Firmani correlation ---}
The Ghirlanda and the Liang \& Zhang correlations share the fact
of using two quantities of the prompt phase ($E_{\rm iso}$ and $E_{\rm p}$),
and one from the afterglow ($t_{\rm j}$).
The Firmani correlation, instead (Firmani et al. 2006a),
links three quantities of the prompt
emission: the peak bolometric and isotropic luminosity $L_{\rm p}$,
the peak energy $E_{\rm p, iso}$ (of the time integrated spectrum), and
a characteristic time: $T_{0.45}$, which is the time interval during
which the prompt emission is above a certain level.
This time was used previously to characterize the variability
properties of the prompt (Reichart et al. 2001).
The correlation, shown in Fig. \ref{f06}
is of the form: $L_{\rm p, iso} \propto E_{\rm p}^{3/2} T_{0.45}^{-1/2}$.
Also this relation is model--independent and assumption--free.
In the comoving frame the luminosity
is a factor $\Gamma^{-2}$ smaller, while the peak energy is $\propto \Gamma^{-1}$
and the time is $\propto \Gamma$.
Therefore the Firmani correlation is ``Lorentz invariant", in the sense
that it has the same form also in the comoving frame.

\subsection{Correlating the correlations}
One can wonder if there are links between these spectral--energy
correlation, highlighting same important GRB physics.
We have already mentioned that the Ghirlanda correlation
is a nice ``explanation" of the phenomenological Liang \& Zhang correlation.
Consider GRBs with the same $E_{\rm p}$ but different $E_{\rm iso}$.
By using $t_{\rm j}$, we obtain that these GRBs have the same $E_\gamma$.
Consider now GRBs that have the same $E_{\rm p}$ but different
$L_{\rm p, iso}$.
The use of $T_{0.45}$ makes them to ``collapse" in the Firmani relation,
which can be thought as the analogous of the Liang \& Zhang relation
for $L_{\rm p, iso}$, instead of $E_{\rm iso}$.

Write the Ghirlanda correlation as $E_\gamma\propto E_{\rm p}^q$,
and compare with the Amati correlation ($E_{\rm iso}\propto E_{\rm p}^2$). 
One obtains $\theta_{\rm j}^2 \propto E_\gamma^{(q-2)/q}$.
If $q=1$ (wind) we have $E_\gamma \propto \theta_{\rm j}^{-2}$, 
while if $q\sim 2/3$ (homogeneous density), 
$E_\gamma \propto \theta_{\rm j}^{-1}$.
If $q= 2$ then the Ghirlanda and Amati correlations are parallel,
and the jet angle distribution is the same for low and high $E_{\rm iso}$.
The dispersion of the Amati correlation can be entirely explained as
the the dispersion of the jet angle, for bursts of the same $E_\gamma$
and $E_{\rm p}$.

The fact that the Amati and the Yonetoku relations are parallel
suggests that the burst duration does not play a crucial role
for defining these two correlations,
even if it does when constructing the Firmani correlation.

\section{Interpretations}

All these correlations are not yet fully understood, but there
were a few suggestions, involving viewing angle effects (Eichler \& Levinson 2006;
Levinson \& Eichler 2005)
or thermal (black--body) emission (Rees \& Meszaros 2005; 
Thompson 2006; Thompson, Meszaros \& Rees 2007) produced not at
the start of the fireball, but during later dissipation of the
kinetic energy of the fireball itself.
With respect to any other process, the black--body has
the simplest link between total emitted energy and peak of 
the spectrum (controlled by the temperature).
Any other emission process would require to specify the density  and/or the magnetic
field, and so on.
The fact that a black--body spectrum is not seen in the time integrated
spectra of bursts may be the result of spectral evolution (the 
temperature may change in time), or the result of an hybrid spectrum
(black--body plus power law, see Ryde 2005) describing the time resolved
spectra.
However, Ghirlanda et al. (2007b) showed that this hybrid model, 
that can fit the 50--1000 keV BATSE spectrum, 
is inconsistent with low energy (2--28 keV) $Beppo$SAX WFC data.

\section{Pretending to do Cosmology}

We would like to use GRBs as standard candles
(or, rather, candles of known luminosity),
but to this aim we have to standardize their 
power/energetics through correlations which
have to be found by adopting a given cosmology.
We have to consider this {\it circularity} problem 
(as well as other problems) before doing cosmology:

\begin{itemize}

\item {\bf Calibration ---}
Usually, to calibrate a cosmology dependent correlation,
one uses a sufficient number of sources at low redshift
($z<0.1$), where the luminosity distance depends weakly on the adopted
$\Omega_{\rm M}$, $\Omega_\Lambda$ values.
For GRBs this is not possible, given the very few of them
observed and predicted at low $z$.
On the other hand, as
Ghirlanda et al. (2006) and Liang \& Zhang (2006) have shown,
to calibrate the correlation it is enough to have a sufficient number of
GRBs in a narrow redshift bin. 
A dozen of GRBs in $\Delta z/z \sim 0.1$ are sufficient.

In the meantime, we have found ways to treat the
circularity problem, based on the scatter of the
used correlation found with different values
of $\Omega_{\rm M}$ and $\Omega_\Lambda$.
The more advanced method is described in Firmani et al. (2005),
and involves a Bayesian--like  approach.

\item {\bf Lensing ---}
The advantage of using GRBs for cosmology is to use high--$z$ objects.
But just because of that, GRBs are exposed to the risk to be biased by gravitational
lensing, affecting their apparent luminosity or energetics.
On the other hand, contrary to other astronomical sources,
GRBs are transient events, making certain types of gravitational
lensing recognizable through the repetition of the light curve
with the same spectrum. 
In other cases, when the afterglow of the GRB is gone, the lensing galaxy may be 
found.
Therefore lensing may not be a problem as serious as it appears at first sight.

\item {\bf Evolution ---}
Like many other class of sources, also GRBs can evolve with cosmic time,
especially because there might be a link between GRBs
and the metallicity of the progenitor star.
But if the spectral energy relations are controlled
by the radiative process, then also $E_{\rm p}$ evolves,
leaving the correlation unaltered.
The isotropic energy $E_{\rm iso}$ may also be affected by the
evolution of the typical jet opening angle.
In this case the Amati and Yonetoku relations may be affected,
but not the Ghirlanda and the Liang \& Zhang correlations.

\item {\bf Outliers ---}
We know that there are at least two outliers to all correlations
(GRB 980425 and 031203). 
There could be more.
This is not ``dangerous" per se, since these two outliers are easily
recognizable, being anomalous in many ways.
The ``danger" comes from GRBs that -- say -- obey the Amati relation
but not the Ghirlanda one, and do not clearly ``stand out" in the
spectral--energy planes.
We claimed (Ghirlanda et al. 2007a) that up to now there are no new
outliers, even including Swift bursts, but this will remain an issue
until many more bursts (with the required information)
will be available.

\end{itemize}

\subsection{Jet breaks: where have they gone?}

In the pre--Swift era the main source of information for the
temporal behavior of the afterglow was the optical.
The achromaticity of the jet break was then checked by
comparing the light curve in different optical filters,
but it was almost never possible to test it with X--ray data.
The latter started 6--9 hours after the trigger, and showed a 
power law decay in time, with no rebrightening or flares
(with rare exceptions). 
The extrapolation of the X--ray lightcurve back in time matched
the end of the prompt emission (once accounting for the different energy bands).
Therefore the Swift observations of the early phases of the X--ray 
afterglow came as a surprise:
there are in fact at least three phases
(Tagliaferri et al. 2005; Nousek et al. 2006): 
a steep initial decay followed by a flat phase
and finally by a steeper decay (similar to what observed in the the
pre--Swift afterglow light curves). 
Often, in addition to this steep--flat--steep behavior, there 
are flares (Burrows et al. 2007), even at relatively late times (hours).
The optical tracks the X--rays sometimes, but more often is different.  
Also in the optical there can be a more than one break, possibly not
simultaneous with X--ray ones (e.g. Panaitescu 2007).

\begin{figure*}[t!]
\begin{tabular}{cc}
\hskip -1 true cm
% \resizebox{\hsize}{!}
% {\includegraphics[clip=true]{firmani_oo.ps}}
\psfig{figure=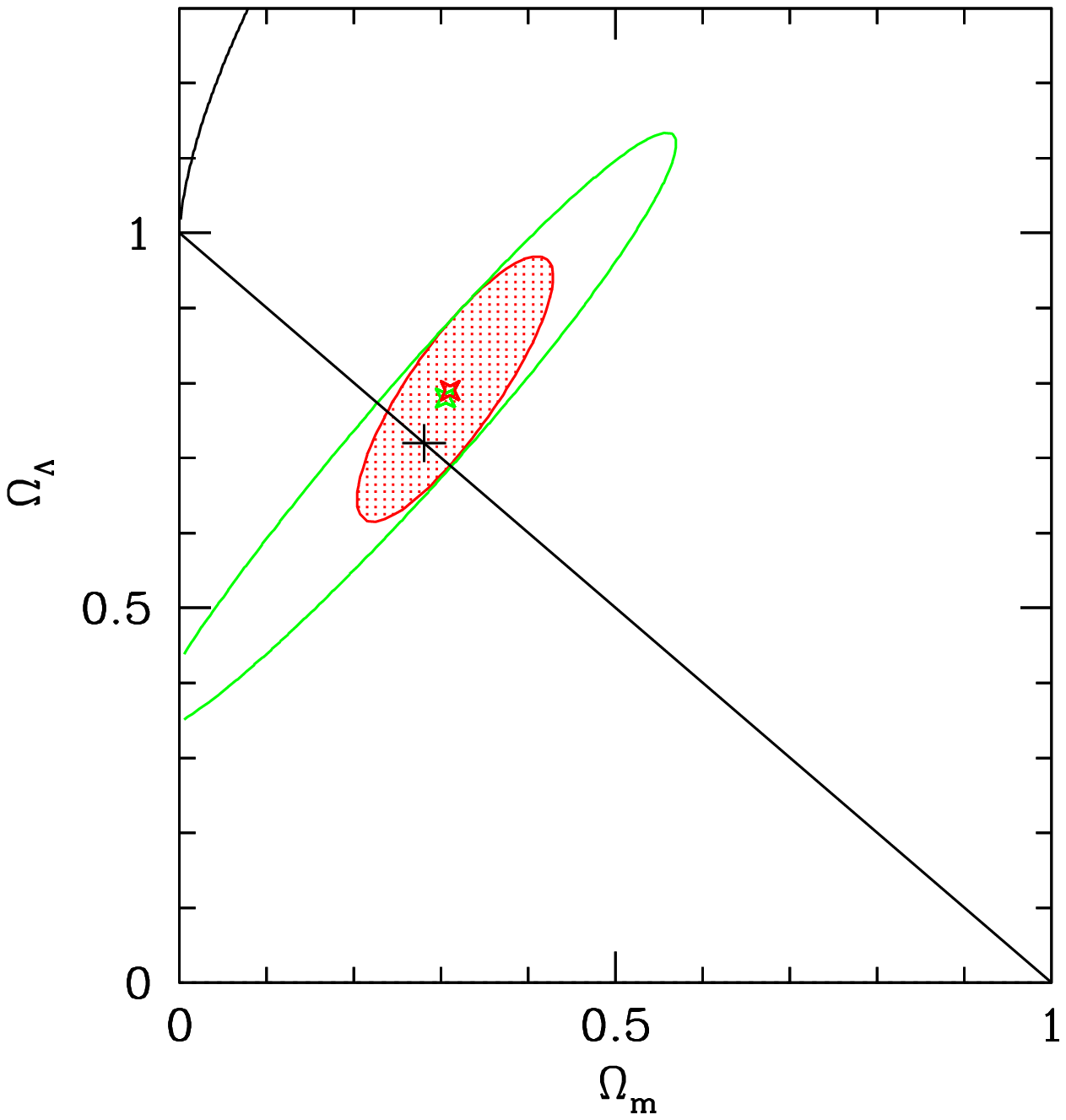,width=8cm,height=8cm}
&\psfig{figure=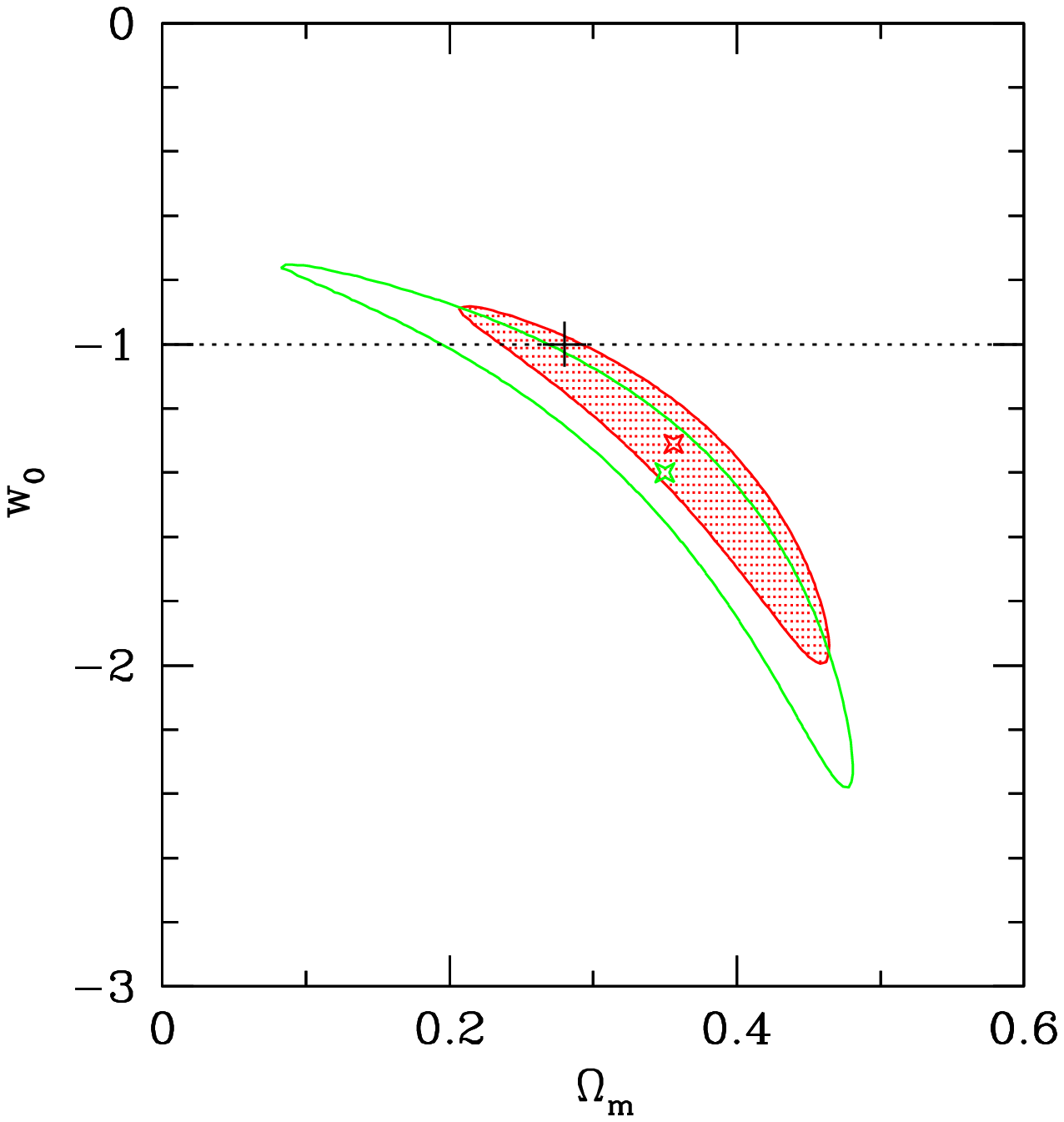,width=8cm,height=8cm}
\end{tabular}
\vskip -1 true cm
\caption{
\footnotesize
Constraints in the $\Omega_\Lambda$--$\Omega_{\rm M}$ plane (left) and
in the $w_0$--$\Omega_{\rm M}$ plane (right).
These constraints have been obtained
using GBRs and SN Ia together.
In both panels, the light grey contours refer to
SN Ia alone. GRBs have been standardized using the Firmani correlation,
SN Ia comes from the sample of Astier et al. (2006).
(From Firmani et al. 2006b).
}
\label{foo}
\end{figure*}

There have been many proposals to explain this unforeseen behavior
in the framework of the external shock model for the afterglow
(see Zhang 2007 for a review), but the fact that the behavior of the
light curve in the optical and X--ray bands seems different 
might suggest that the two components come from different 
emitting regions (Uhm \& Beloborodov 2007; Grenet, Daigne \& Mochkovitch 2007;
Ghisellini et al. 2007).

All the above implies that we must be careful when identifying a break with 
the jet break time.
Since in the pre--Swift era optical break times
were used, it is safer to use {\it only} the optical light curves to find
$t_{\rm j}$, and relax the requirement that the break should be present also
in the X--ray light curve which could be produced by a different mechanism.

\section{First results}

There have been several attempts to use GRBs to constrain
the cosmological parameters, by our group as well as by others.
The found results are encouraging and very similar independently on
the method used to standardize their energetics/peak luminosity.
In other words, using the Ghirlanda, Liang \& Zhang or the Firmani correlations
gives consistent results.
GRBs alone cannot (yet) compete with SN Ia, given the small number
of GRBs with known redshift, $E_{\rm p}$ and $t_{\rm j}$ or $T_{0.45}$.
Furthermore, the uncertainties associated with GRBs are larger than for SN Ia,
but this is partly compensated by the larger redshifts of GRBs.
% For illustration, we show in
% Fig. NN the contours in the $\Omega_\Lambda$--$\Omega_{\rm M}$ 
% plane using the 25 GRBs defining the most updated Ghirlanda relation 
% (Ghirlanda 2007c):  going from 18 to 25 sources
% makes the contours to shrink notably.

In Fig. \ref{foo} we show the constraints in the 
$\Omega_\Lambda$--$\Omega_{\rm M}$ plane obtained by Firmani et al. (2006b)
using a sample of 19 GRBs and the sample of SN Ia of Astier et al. (2006).
The peak luminosity of GRBs is standardized using the Firmani correlation.
The figure also reports the contours obtained SN Ia only.
As can be seen, despite the still very small sample of available GRBs,
the contours of the combined sample are remarkably smaller.
The concordance cosmology ($\Omega_\Lambda\sim 0.7$, $\Omega_{\rm M}\sim 0.3$)
is confirmed.

GRBs, together with SN Ia, can be also used to put constraints
on the equation of state of Dark Energy:
\begin{equation}
P(z) \, = \, w(z) \rho c^2
\end{equation}
where $P$ is the pressure and $\rho c^2$ the energy density of the Dark Energy. 
The case of a cosmological constant corresponds to $w(z)=-1$,
and different models are described by different $w(z)$ laws.
The right panel of Fig. \ref{foo} shows the constraints using GRBs+SN Ia in the plane
$w_0$--$\Omega_{\rm M}$, where it is assumed that 
$w_0=w(z)$ is a constant, but can be different from --1.

\section{Conclusions}

Gamma Ray bursts can be a novel class of standard candles.
Potentially, they are detectable at any redshift,
and their prompt emission is free from extinction problems.
On the other hand, we still need to understand the physical reasons
of the found spectra--energy correlations, check for evolution--induced
effects and the possible existence of outliers.
GRBs should be thought as complementary to SN Ia, able to measure
the Universe in a redshift range that SN Ia cannot reach.

\begin{acknowledgements}
I am grateful to G. Ghirlanda, A. Celotti, C. Firmani, D. Lazzati,
F. Tavecchio, L. Nava, M. Nardini and E. Lisiero for fruitful discussions.
\end{acknowledgements}

\bibliographystyle{aa}

\end{document}